\documentclass[twocolumn,showpacs,pre]{revtex4}

\usepackage{graphicx}
\usepackage{dcolumn}
\usepackage{bm}

\newcommand{\di}{{d}}
\newcommand{\kT}{k_{\rm B}T}
\newcommand{\angw}[1]{\left<#1\right>_\omega}
\newcommand{\mref}[1]{(\ref{#1})}

\begin{document}

\title{A possible new phase of antagonistic nematogens in a disorienting field}

\author{T. G. Sokolovska$^{1,2}$}
 \email{tata@icmp.lviv.ua}

\author{M. E. Cates$^1$}%

\author{R. O. Sokolovskii$^{1,2}$}

\affiliation{$^1$School of Physics, JCMB King's Buildings, University of Edinburgh, Edinburgh EH9 3JZ, United Kingdom\\
$^2$Institute for Condensed Matter Physics, 1 Svientsitskii, Lviv 79011, Ukraine}

\date{\today}

\begin{abstract}
A simple model is proposed for nematogenic molecules that favor perpendicular orientations as well as parallel ones. (Charged rods, for example, show this antagonistic tendency.) When a small disorienting field is applied along $z$, a low density phase $N_-$ of nematic order parameter $S_z<0$ coexists with a dense biaxial nematic $N_b$. (At zero field, $N_-$ becomes isotropic and $N_b$ uniaxial.) But at stronger fields, a new phase $N_{+4}$, invariant under $\pi/2$ rotations around the field axis, appears in between $N_-$ and $N_b$. Prospects for finding the $N_{+4}$ phase experimentally are briefly discussed. 
\end{abstract}


\maketitle

Practical applications of nematics are based on their strong responses to weak external influences. Therefore, the phase behavior of nematics in external fields of different symmetries is a subject of many investigations (see \cite{Fan,VargaPalfySokolovska} and references therein). It was shown that external fields can lead to major changes in phase transition behavior. 
In nematics with positive anisotropy, an external magnetic or electric field is `orienting': it favors positive nematic order parameter $S$. 
In strong orienting fields the first order phase transition between the `isotropic' phase (which for nonzero field acquires finite $S>0$) and the nematic phase $N$ is suppressed above a critical temperature. In systems with negative molecular anisotropy, similar fields have instead a disorienting effect, favoring $S<0$. The isotropic phase then acquires uniaxial planar orientation; with decreasing temperature or increasing density, this phase $N_-$ has a phase transition into a biaxial phase $N_b$ (which connects smoothly to $N$ as the field is removed). A critical field is expected, above which the $N_-/N_b$ phase transition becomes second order \cite{Fan}. 

Previous theoretical investigations of external field effects (in cases where the interaction potential was explicitly specified \cite{hess}) were carried out \cite{VargaPalfySokolovska} for molecular potentials proportional to the second Legendre polynomial of the relative molecular orientation, $P_2(\cos\omega_{12})$, which is the simplest dependence favouring parallel alignment of nematic particles. In practice, though, real molecules have interactions described by more complex orientational dependences than this. For example, uniaxial nematogenic particles may display disposition towards mutually perpendicular orientations as well the parallel one. The molecular interaction then has minima at both $\omega_{12} = 0$ and $\omega_{12} = \pi/2$. (Generically these are of unequal depths.) An example of this antagonistic tendency is for the case of charged rods. It was shown that the Debye-Huckel expression for the electrostatic repulsion between two effective line charges at mutual angle $\omega_{12}$ is proportional to $|\sin\omega_{12}|^{-1}$. As a consequence, two charged rods tend to rotate to a perpendicular configuration; this interaction is antagonistic to the usual (steric) Onsager interaction, favoring parallel alignment, which is also present. Such an electrostatic twisting results in significant quantitative changes of the nematic phase diagram in zero field (see Ref.~\cite{Vroege} and references therein). Some nonconvex particles, dumbbells for example, should have a certain tendency to mutual twisting too.

The simplest angular dependence of a potential describing molecules with a tendency to orient perpendicularly as well as parallel to each other is the fourth order Legendre polynomial
\begin{equation}
P_4(\cos\omega_{12})=[35\cos(4\omega_{12})+20\cos(2\omega_{12})+9]/64
\label{p4}
\end{equation}
In general, this $P_4$ interaction should be taken in addition to, not in place of, the customary $P_2(\cos\omega_{12})$ dependence. However, absence of a $P_2$ term does not imply higher molecular symmetry than when it is present; a pure $P_4$ interaction already has unequal well depths for perpendicular and parallel interactions. 

The task of this paper is to consider an `antagonistic' nematic in the presence of an external disorienting field. The origin of the disorienting field may vary: electric or magnetic fields, orienting flow in side-chain polymers (with a side group trend to be perpendicular to the backbone), surface influences and so on. The only important factor is that the particles want to orient perpendicularly to the field direction.
Note that besides being ubiquitous in display applications, which involve both surface and electric fields \cite{Shashidhar}, the disorienting effect in a sample could be strongly enhanced in some cases by addition of impurities \cite{Swager}.

Our model comprises a system of uniaxial molecules in a disorienting field of strength $W$, where the potential of the molecular interaction with the field is 
\begin{equation} 
v(1)=v(\omega_1)=W(3\cos^2\theta_1-1)/2, \;\;      W>0,
\end{equation}
$\omega_i=(\theta_i,\varphi_i)$ being the orientation of molecule $i$. The above formula assumes that the field is directed along the $z$ axis and states that molecular orientations along the field direction are energetically unfavorable. The pair potential is taken to be a sum of the hard sphere potential for spheres of diameter $\sigma$, and of an anisotropic part
\begin{equation}
v(1,2)=\sum_{n=2,4}v_n(R_{12})P_n(\cos\omega_{12})
\label{V1},
\end{equation}
where $\omega_{12}$ is the angle between the axes of molecules $1$ and $2$; $R_{12}$ is the distance between their centers; and
\begin{equation}
v_n(R_{12})=-A_n(z_n\sigma)^2\frac{\exp(-z_nR_{12})}{R_{12}/\sigma}
\label{V2}
.
\end{equation}
In the limit $z_n\sigma\to 0$, this model belongs to a family of the so-called Kac potentials, for which the mean field approach is accurate, and the free energy obeys:
\begin{eqnarray}
\beta F/N&=& F_0+\int f(\omega)[\ln f(\omega)-1]\di\omega
\nonumber \\ 
&+&\beta W\int f(\omega)P_2(\cos\theta)\di\omega
\\ \nonumber
&-&\sum_{n=2,4}12\beta A_n\eta \int f(\omega_1)P_n(\cos\omega_{12})f(\omega_2)\di\omega_1\di\omega_2
,
\end{eqnarray}
where $\eta=\pi(N/V)\sigma^3/6$, $f(\omega)$ is a single particle distribution function, and for the free energy of the reference hard sphere system we use the ``quasiexact'' Carnahan-Starling expression,
$
F_0=\ln\eta +(4\eta-3\eta^2)/(1-\eta)^2+ \mbox{const}
$.

To make the mathematics more compact, we give details only for the particular case $A_2=0$. (However, adding a $P_2(\cos\omega_{12})$ term is straightforward.) Using a spherical harmonic expansion for $P_n(\cos\omega_{12})$ \cite{Gubbins} and noting that the model under consideration is non-polar, one can get the explicit expression for $\Delta F\equiv\beta F/N-F_0$:
\begin{eqnarray}
\Delta F&=& \int f(\omega)[\ln f(\omega)-1]\di\omega
+\beta W\angw{P_2(\cos\theta)}
\label{F}\\ \nonumber
&-&\beta A_4\eta \frac{12}9[
\angw{Y_{40}(\omega)}^2
+2\angw{Y_{42}^c(\omega)}^2
+2\angw{Y_{44}^c(\omega)}^2
],
\end{eqnarray}
where $\angw{\cdots}=\int f(\omega)(\cdots)\di\omega$ denotes orientational averages within a self-consistent one-particle distribution function. The superscript `\textit{c}' means that the real part of the spherical harmonic $Y_{ij}$ is taken; \mref{F} is written assuming that the $zx$-plane is a symmetry plane. This overcomes a continuous degeneracy of the free energy connected with the fact that a spontaneous ordering can rotate without any energy cost in the $xy$-plane. 

Minimizing \mref{F} with respect to $f(\omega)$ we obtain the self-consistent single particle distribution function (with $C$ a normalization constant)
\begin{eqnarray}
f(\omega)&=&C\exp\{-\beta W\angw{P_2(\cos\theta)}
\nonumber\\&&
+\frac{24}9\beta A_4\eta [
\angw{Y_{40}(\omega)}Y_{40}(\omega)
\label{ff}
\\ \nonumber &&
+2\angw{Y^c_{42}(\omega)}Y^c_{42}(\omega)
+2\angw{Y^c_{44}(\omega)}Y^c_{44}(\omega)
]
\}.
\end{eqnarray}
We calculated the phase diagrams of this system using, where appropriate, the (standard) common-tangent construction to find coexisting states of different density. The symmetries of all the resulting phases can be defined using three order parameters:
\begin{eqnarray}
S_z&=&\angw{P_2(\cos\theta)}
,\\
S_x&=&\angw{\sin^2\theta\cos(2\varphi)}
,\\
S_{xy}&=&\angw{\sin^2\theta\cos(4\varphi)}
.
\end{eqnarray}
Here $S_z$ and $S_x$ describe alignment of molecules along the corresponding axes, whereas $S_{xy}$ describes ordering along the $y$ axis as well as along the $x$ axis; when $S_x=0$, $S_{xy}$ describes ordering with fourfold symmetry.

\begin{figure}
\includegraphics[width=85mm,clip=true]{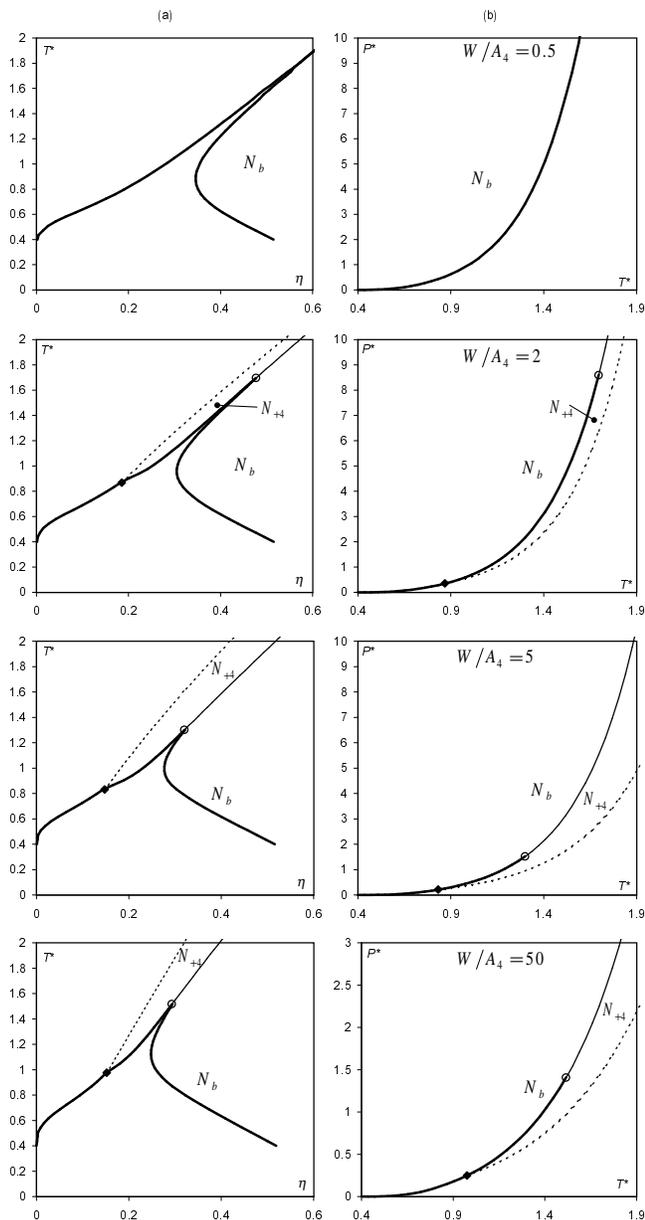}
\caption{\label{fig:pd}The phase diagram of the system ($A_2=0$) in (a) the temperature-density
($T^*=\kT/A_4$, $\eta=\pi\rho\sigma^3/6$) and (b) pressure-temperature coordinates ($P^*=(\pi\sigma^3/6)(P/A_4)$) for different values of the disorienting field ($W/A_4$). $N_{+4}$ is the phase with $S_{xy}\ne0$, $S_x=0$, $N_b$ denotes the biaxial nematic phase. Dashed lines correspond to the the second order phase transition into $N_{+4}$ phase, thin solid lines likewise into $N_b$; thick lines denote the first order transition. Circles are tricritical points, diamonds denote critical end points.
}
\end{figure}

In Fig.~\ref{fig:pd} the phase diagrams are presented for different values of the ratio $W/A_4$ describing the external field strength in units of the molecular interaction. (We set $A_2=0$ as explained above.) At zero field ($W/A_4\to - 0$) the system undergoes a first order phase transition from an isotropic to a uniaxial nematic $N$; this phase has $S_z>0$, $S_x=S_{xy}=0$. Small disorienting fields ($W/A_4=0.5$) transform the system symmetry, so that the isotropic phase at low densities is replaced by a uniaxial planar phase $N_-$ (this has $S_z<0$, $S_x=S_{xy}=0$). At intermediate densities (between the bold solid lines in the figure) this coexists with a highly ordered biaxial nematic phase $N_b$. The biaxial has $S_z<0$, $S_x>0$, $S_{xy}>0$, and becomes a stable single phase at high densities. At higher fields the orientational transition becomes second order at high temperatures, joining onto the first order transition (which is accompanied by the miscibility gap) at the tricritical point. The wide miscibility gap at low temperatures is connected with the appearance of an effective attraction between oriented molecules, which causes the gas liquid phase transition. Molecular potentials of typical thermotropic nematics are characterized by strong isotropic attractions, in addition to anisotropic interactions. In this case the condensation could arise in a non-ordered phase. However, some lyotropic systems display a wide miscibility gap between phases of different symmetries. So far, the picture is qualitatively just as found previously for a pure $P_2(\cos\omega_{12})$ interaction \cite{Fan}. At low disorienting field strengths, then, the antagonistic tendency of our molecules cannot manifest itself and the behavior is that of a conventional nematic system. 

However, at strong enough fields (e.g., $W/A_4=2$) the situation differs strongly from the conventional case: two consecutive orientational transitions now take place. The first one is always of second order and meets the binodal at a critical end point. The resulting phase is invariant under rotations by $\pi/2$ around the field direction; it has $S_z<0$, $S_x=0$, and $S_{xy}>0$. In accordance with the general classification \cite{Toner}, this new phase should be denoted as $N_{+4}$. The second orientational phase transition is from $N_{+4}$ into $N_b$; this changes from first to second order at the tricritical point as before. The new $N_{+4}$ phase is (like $N_b$) a result of spontaneous ordering, and is characterized  by long-range correlations of Goldstone type. Note that the molecular potential under investigation does not possess the symmetry of the $N_{+4}$ phase (even for the particular case studied, of $A_2=0$). In some respects the situation is similar to that found by computer simulations, when a cubatic state appears in a system of hard truncated spheres \cite{Frenkel}. The $N_{+4}$ phase is not cubatic but might be called `square-atic', since it has fourfold rotational order in the transverse plane only.

From the calculated phase diagrams (especially in $T-\eta$ coordinates) one can see a nonmonotonic efect of the external field on the $N_{+4}$ region. The widest $N_{+4}$ region arises at intermediate field values. We attribute this fact to the disorienting field's ability to restrict the part of orientational phase space in which particles mainly reside. Indeed, in the limit of an infinite field, the particle orientations are confined in the plane perpendicular to the field. In this case the system can be described by the plane rotator model \cite{Romano}; this has the same phase diagram (independent of $n$) for all pair potentials proportional to $\cos(n\omega_{12})$. In contrast, with an unrestricted orientational phase space (that is, at zero field), the temperature at which spontaneous ordering occurs decreases with increasing $n$. Disorienting fields (pushing the system towards the plane rotator state) promote the ordering with $S_{xy}\ne 0$ more strongly than that with $S_x\ne 0$, so that the $N_{+4}$ phase grows leaving behind competing ordered phases such as $N_b$. However, in strong enough fields the susceptibility of the order parameter $S_{xy}$ to the field decreases, whereas the $N_b$ ordering is not saturated. As a result, in strong enough fields the $N_b$ phase begins to displace the $N_{+4}$ region.

This field effect on the $N_{+4}$ region is closely linked with the tricritical behavior (Table~\ref{tab:1}). The nonmonotonic field effect on where the tricritical point lies can be explained by the existence of two competing tendencies \cite{PRL}. First, the external field furthers the orientational ordering of particles and, therefore, causes a more effective attraction between particles. This favors the phase separation and raises the `binodal' in temperature. The second tendency takes place if the susceptibility of the low-density phase ($N_{+4}$) to the field is larger than that of the coexisting dense ($N_b$) phase. This decreases the energy gain of the phase separation. The second tendency is very strong for our system. Only at very strong fields ($W/A_4=50$) do the susceptibilities of the coexisting phases become comparable, allowing the first tendency to win at last.

\begin{table}
\caption{\label{tab:1} Coordinates of the tricritical point ($A_2=0$) at different values $W/A_4$ of the disorienting field. (Notation as in Fig.~\ref{fig:pd}.)}
\begin{ruledtabular}
\begin{tabular}{cccc}
$W/A_4$ & $T^*$ & $\eta$ & $P^*$\\
\hline
1       & 2.02  &  0.62  & 39.1 \\
2       & 1.70  &  0.48  & 8.59 \\
5       & 1.30  &  0.32  & 1.52 \\
50      & 1.52  &  0.29  & 1.40 \\
\end{tabular}
\end{ruledtabular}
\end{table}

\begin{figure}
\includegraphics[width=85mm,clip=true]{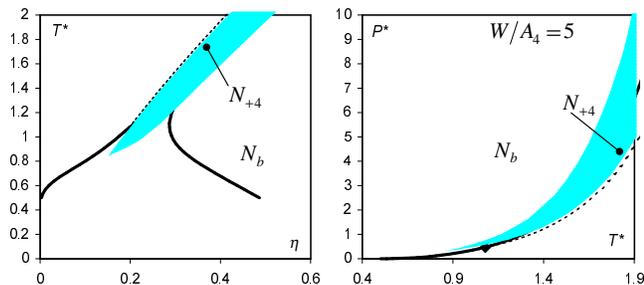}
\caption{\label{fig:pd2}The phase diagram of the model \mref{V1} with $A_2=0.1A_4$ in the temperature-density 
and pressure-temperature coordinates 
for a moderate disorienting field ($W/A_4=5$). 
(Notation as in Fig.~\ref{fig:pd}.)
For comparison, the location of the $N_{+4}$ phase for the case $A_2=0$ is shaded.
}
\end{figure}

To study the robustness of the predicted $N_{+4}$ phase with respect to changes of the intermolecular potential we calculated phase diagrams for $A_2/A_4=0.1$; $0.2$; $0.3$ at $W/A_4=5$; 10; 50; $\infty$ in a density interval sensible for liquids ($\eta<0.6$). In Fig.~\ref{fig:pd2} the phase diagram for $A_2=0.1A_4$ is presented (setting $W/A_4 =5$). With this potential, the parallel orientation of molecules is more energetically favorable than in the case $A_2=0$. As a consequence, the $N_b$ transition temperature increases, whereas the temperature of the $N_{+4}$ transition is not affected significantly. The $N_{+4}$ region is reduced in size, but the phase remains stable. The phase transition $N_{+4}/N_b$ is first order now (where previously it was second order, for this value of $W/A_4$) which should lead to a jump in the elastic constants, for example. For all the investigated cases with $A_2>0$ the phase transition into $N_b$ is first order (even at infinite field) whereas the $N_-/N_{+4}$  transition remains second order. It follows from our results that the $N_{+4}$ region is broader in density and temperature at the moderate field $W/A_4=10$. Moreover, for $A_2=0.2A_4$ (when the parallel orientation of molecules is 4.4 times more energetically favorable than the perpendicular one) the $N_{+4}$ phase exists at $W/A_4=5$; 10; 50, but is absent in infinite and zero fields. When the parallel orientation is almost 6 times more favorable than the perpendicular one ($A_2=0.3A_4$) the $N_b$ phase finally excludes $N_{+4}$.

In conclusion, we have studied a model system of uniaxial particles with antagonistic molecular interactions in the presence of a disorienting field causing an easy-plane anisotropy. This may lead to spontaneous orientational ordering of two types. The first is a biaxial nematic $N_b$ with a nonzero value of the conventional biaxial order parameter $S_x=\angw{\sin^2\theta\cos(2\varphi)}$ (as well as of the uniaxial order parameter $S_z=\angw{P_2(\cos\theta)}$). The biaxial order parameters is however zero in the second spontaneously ordered phase, named $N_{+4}$, which we predict to arise at intermediate density. The symmetry of this phase is defined instead by a nonvanishing higher orientational average $S_{xy} = \angw{\sin^2\theta\cos(4\varphi)}$. This corresponds to a cross-shaped distribution of preferred molecular orientations, which can be rotated in the $xy$-plane without any energy cost (just like the usual transverse director in the $N_b$ phase). This feature is connected with Goldstone-mode fluctuations of particle orientations in the plane. 

Device applications for nematic phases are of widespread importance \cite{Shashidhar}, and a number have already been envisaged for the biaxial nematic phase \cite{Luckhurst}. But thermotropic biaxial nematics have proved to be very elusive. One of the causes is that they are displaced by other structures. It makes it necessary to study all phases nearby (like $N_{+4}$) and conditions of their existence. An important question that arises is how to look for the new $N_{+4}$ phase experimentally. $N_{+4}$ could be confused with $N_b$ in its elastic behavior. Both should demonstrate elastic resistance to the following distortions: 1) splay and bend in the $xy$-plane, 2) twist in the $z$ direction. In contrast to $N_b$ and  $N_{+4}$, the uniaxial phase $N_-$ does not posess such elastic behavior. The new symmetry of the $N_{+4}$ phase also should influence the behavior of defects, as is also true for biaxial nematics \cite{UFN}. On the other hand, since it has a fourfold symmetry axis, the  $N_{+4}$ phase is uniaxial with respect to second rank tensors, and therefore not distinguishable by simple dielectric or optical measurements from an ordinary uniaxial nematic.

Finally it is worthwile to note that the $N_{+4}$ symmetry could in principle appear without external fields, in molecules of a more complex geometry. From this point of view the investigation of systems of biaxial particles in search of the $N_{+4}$ phase represents an interesting prospect. The ordering of long molecular axes eventually leads to the restriction of orientational phase space of the molecular short axes without any external field. This can influence their further orientational ordering with the same consequences as discussed above.

R.~O.~S. thanks NATO and the Royal Society for financial support.

\end{document}